\documentclass[superscriptaddress,showpacs,
amsmath,amssymb,
prc
floatfix, twocolumn ]%
{revtex4-1}
\usepackage{color}
\usepackage{newtxtext}
\usepackage{newtxmath}
\usepackage{multirow}
\definecolor{mygray}{gray}{.9}
\definecolor{mypink}{rgb}{.99,.91,.95}
\definecolor{mycyan}{cmyk}{.3,0,0,0}
\usepackage{graphicx}
\usepackage{dcolumn}
\usepackage{bm}
\usepackage{booktabs}
\usepackage{longtable}

\begin{document}
	
\title{Study of neutron-deficient nucleus $^{224}$Np and its $\alpha$ decay by particle-number-conserving method in the framework of deformed shell model }
	
\author{Xiao-Tao He}%
\email{hext@nuaa.edu.cn}
\affiliation{College of Materials Science and Technology, Nanjing University of Aeronautics and Astronautics, Nanjing 210016, China}
\author{Kun Huang}%
\affiliation{College of Materials Science and Technology, Nanjing University of Aeronautics and Astronautics, Nanjing 210016, China}
\author{T. M. Shneidman}%
\affiliation{Bogoliubov Laboratory of Theoretical Physics, Joint Institute for Nuclear Research, Dubna 141980, Russia}
\author{N. V. Antonenko}%
\affiliation{Bogoliubov Laboratory of Theoretical Physics, Joint Institute for Nuclear Research, Dubna 141980, Russia}
\author{Jun Zhang}%
\affiliation{College of Physics, Nanjing University of Aeronautics and Astronautics, Nanjing 210016, China}
\author{Hong-qiang You}%
\affiliation{College of Physics, Nanjing University of Aeronautics and Astronautics, Nanjing 210016, China}

\date{\today}
	
\begin{abstract}     
    The particle-number-conserving (PNC) method in the framework of the deformed shell model (DSM) is employed to study the properties of the newly discovered short-lived neutron-deficient nucleus $^{224}$Np and its $\alpha$-decay. The calculated energy of $\alpha$-particle lies within 300 keV of the experimental data. This is the first application of the PNC method to the region of neutron-deficient nuclei. This work provides the first attempt to combine the microscopic PNC theory with empirical formulas to study the nuclear $\alpha$ decay. The configurations of ground states are assigned as $ \pi 5/2^{-}[523]\otimes\nu 5/2^{+}[633] $ for $ ^{224} $Np, $ \pi 1/2^{-}[530]\otimes\nu 3/2^{+}[642] $ for $ ^{220}$Pa,  $ \pi 3/2^{+}[651]\otimes\nu 1/2^{-}[501] $ for $^{216} $Ac, and $ \pi 7/2^{-}[514]\otimes\nu 3/2^{-}[501] $ for $ ^{212} $Fr. The absence of the $Z=92$ subshell closure in $^{224}$Np is explained by analyzing the proton single-particle levels. Low-lying excited states are predicted for nuclei along the $\alpha$-decay chain by the PNC method. Based on the PNC predicted $\alpha$-decay energy and the assigned configurations, the $\alpha$-decay half-lives are calculated by the empirical formulas, in which the angular momentum taken away by the $\alpha$ particle is taken into account. The angular momentum have an important effect on the $\alpha$-decay half-life. The errors of the empirical formulas calculation are in two orders of magnitude with the experimental data.   	
\end{abstract}
	
\maketitle
\section{Introduction}
    Radioactive ion beams have greatly expanded our capabilities in the study of nuclear physics, in particular in the study of the structure and properties of nuclei far from the $\beta$ stable line. Among these, neutron-deficient nuclei are of special interest due to their unique nuclear structure, which is crucial to understanding fundamental nuclear interactions~\cite{QiC2016_RiP1_77,BenhamoudaN2004_APHNSIP19_191}. The evolution of the shell structure beyond the heaviest known double-magic nucleus $^{208}$Pb is of great importance for understanding the stability and structure of superheavy nuclei. In this respect, it is interesting to investigate the shell structure around $Z=92$. The $Z=92$ shell gap is predicted in the theoretical  calculations~\cite{RutzK1998_NPA634_67,GengL2005_Potp113_785,GengL2006_23_1139,SagawaH2014_PiPaNP76_76,MoellerP1997_ADaNDT66_131}. This is at variance with large-scale shell model calculations~\cite{CaurierE2003_PRC67_54310}. Experimental results on the $\alpha$-decay and proton separation energies of $^{223}$Np~\cite{Sun2017} and $^{219}$Np~\cite{Yang2018} show no indication of the shell or subshell closure at $Z$~=~92.
    
    New heavy neutron-deficient nuclei have been produced, including $ ^{214,216,218,219,221} $U~\cite{ZhangZ2021_PRL126_152502,MaL2015_PRC91_51302,LeppaenenA2007_PRC75_54307,AndreyevA1993_ZfPAHaN345_247,KhuyagbaatarJ2015_PRL115_242502}, $ ^{220} $Pa~\cite{Ma2021}, and $ ^{219,220,222-224} $Np~\cite{Sun2017,Huang2018,Yang2018,Zhang2019a,Ma2020}.  Physicists have described the neutron-deficient region via different models~\cite{AdamianG2008_PRC78_44605,HeC2022_PRC106_64310}. They drawn the $\alpha$-decay systematics for the isotopes near $N=126$ with $89\leq Z\leq93$, which support that the robustness of $N=126$ shell closure up to the Np isotopes~\cite{Sun2017,Yang2018}. However, research on the $\alpha$-decay of the odd-odd nucleus $ ^{224} $Np remains unknown area in theoretical calculations, lacking any microscopic structural insights. Remarkably, it stands as the heaviest nucleus residing on the proton drip-line to date. 

    For unstable nuclei, especially neutron-deficient nuclei, $\alpha$-decay is a dominant decay mode and plays a significant role in the identification of new elements or new isotopes ~\cite{ZhangZ2021_PRL126_152502,MaL2015_PRC91_51302,LeppaenenA2007_PRC75_54307,AndreyevA1993_ZfPAHaN345_247,KhuyagbaatarJ2015_PRL115_242502,Ma2021,Sun2017,Huang2018,Yang2018,Zhang2019a,Ma2020}. $\alpha$-decay provides information on half-lives, decay energies, spin and parities of ground and excited states, as well as important insight on properties of nuclear interaction and shell effects~\cite{Yahya2022,AkrawyD2019_PRC100_44618,Deng2020,Ren2012,Royer2010,Ren2004,Ni2008,Saxena2021,Zanganah2020,DaeiAtaollah2018}.  In the theoretical study, $\alpha $ radioactivity was explained successfully as a typical quantum tunneling effect by Gamow~\cite{GamowG1928_N122_805} and by Condon and Gurney~\cite{GurneyR1928_N122_439} in 1928. Since then, all kinds of theoretical models and empirical formulas have been proposed~\cite{Poenaru2011,Poenaru2012,Samanta2007,Bhattacharya2007,Basu2005,Chowdhury2006,Chowdhury2008,Bao2015,ZhangH2006_PRC74_17304,Xu2005,Xu2006,BuckB1990_PRL65_2975,RoyerG2002_NPA699_479,Royer2004,Dong2009,Wang2014}. It was found that the energy of $\alpha$-decay plays a crucial role in the description of the decay process. Therefore, investigating the $\alpha$-decay energy is highly significant in revealing nuclear structure and properties of $\alpha$-decay.
   
    In this work, based on the deformed shell model(DSM) with pairing treated by a particle-number-conserving (PNC) method, 
    the $\alpha$-decay energies and the microscopic structure of the low-lying excited and ground-states of the nuclei in the $\alpha$-decay chain of $^{224}$Np are studied. This is the first application of the PNC method to the region of neutron-deficient nuclei and to provide a microscopic description of the nuclear $\alpha$ decay. Using the PNC results of the $\alpha$-decay energies and angular momentum, the half-lives of nuclei along the $^{224}$Np $\alpha$-decay chain are studied by the empirical approaches of Refs.~\cite{Deng2020,Ni2008,Akrawy2019a}. 
	
    This article is organized as follows. In Section~\ref{subsec:PNC}, we present the theoretical framework of the PNC method. The method of calculation of $\alpha$-decay half-lives presented in Section~\ref{subsec:formulas-alpha-decay}. The calculation details and discussion are given in Section \ref{Sec:results}. Section \ref{Sec:sum} is a summary.

\section{Theoretical framework}
\subsection{PNC method}
{\label{subsec:PNC}}
     The detailed description of PNC method can be found in Refs.~\cite{He2018,He2020a,Zhang2023}. Here we give a brief introduction of the related formalism. The deformed shell model Hamiltonian with monopole- and quadrupole- pairing is
	\begin{eqnarray}
		\label{eq:H_CSM}
		H_\mathrm{DSM}=H_{\rm Nil} + H_\mathrm{P}(0)+H_\mathrm{P}(2),
	\end{eqnarray}
     $H_{\rm Nil}=\sum{h_\mathrm{Nil}}$ is the Nilsson Hamiltonian with 
	\begin{eqnarray}
	    \label{eq:h_Nil}
	    h_\mathrm{Nil}&=&\frac{1}{2}\hbar\omega_{0}(\varepsilon_2,\varepsilon_4,\varepsilon_6)\Big[-\bigtriangledown_\rho^2+\frac{1}{3}\varepsilon_2\Big(\frac{2\partial^2}{\partial\zeta^2}-\frac{\partial^2}{\partial\xi^2}-\frac{\partial^2}{\partial\eta^2}\Big)\notag\\
      &+&\rho^2-\frac{2}{3}\varepsilon_2\rho^2P_2(\mathrm{cos}\theta_t)+2\varepsilon_4\rho^2P_4(\mathrm{cos}\theta_t)\\   &+&2\varepsilon_6\rho^2P_6(\mathrm{cos}\theta_t)\Big]-2\kappa\hbar\mathring{\omega}_0\big[\vec{l_t}\cdotp\vec{s}-\mu(\vec{l_t}^2-\langle\vec{l_t}^2\rangle_N)\big]\notag.	
         \end{eqnarray}
    where $\varepsilon_2$, $\varepsilon_4$ and $\varepsilon_6$ are the quadrupole, hexadecapole and high-order deformation parameters, respectively, and the subscript $t$ means that the single-particle Hamiltonian $h_\mathrm{Nil}$ is written in the stretched coordinates $(\xi,\eta,\zeta)$, and $\rho^2=\xi^2+\eta^2+\zeta^2$~\cite{NilssonS1969_NPA131_1}. $H_\mathrm{P}(0)$ and $H_\mathrm{P}(2)$ are the monopole- and quadrupole-pairing correlations, respectively. 

    The $H_\mathrm{DSM}$ is diagonalized in a sufficiently large many-particle configuration (MPC) space, which is constructed by including the configurations with energy $E_i\leq~E_0+E_c$, where $E_0$ is the energy of the lowest configuration and $E_c$ is the cutoff energy. The eigenstate of $H_\mathrm{DSM}$ reads
	\begin{equation}\label{eq:eigenstate}
		| \psi \rangle = \sum_{i} C_i | i \rangle, ~~ C_i \text{ is real.}
	\end{equation}
    $ |i \rangle $ is a configuration of the many-particle system. For two unpaired single particles in an odd-odd nucleus with the seniority $\nu=2$, each $|i\rangle$ in Eq.~(\ref{eq:eigenstate}) has the form 
	\begin{equation}   
	\label{eq:CMPC}
        \begin{aligned}                      |i\rangle&=|\sigma_1\sigma_2\mu_{1}\overline{\mu}_1\cdots\mu_{k}\overline{\mu}_k \rangle \\ &=b_{\sigma_{1}}^{\dagger}b_{\sigma_{2}}^{\dagger}b_{\mu_{1}}^{\dagger}b_{\overline{\mu}_{1}}^{\dagger}\cdots b_{\mu_{k}}^{\dagger}b_{\overline{\mu}_{k}}^{\dagger}|0\rangle,(\sigma\not=\mu),
        \end{aligned}
	\end{equation}
       where $\sigma_1$,~$\sigma_2$ are the two blocked single-particle states. The angular momentum projection along the nuclear symmetry $z$ axis is $K=|\Omega_{\sigma_1}\pm\Omega_{\sigma_2}|$, and the parity is $\pi=(-)^{N_{\sigma_1}+N_{\sigma_2}}$. The configuration of higher-seniority $\nu>2$ states is constructed similarly, and the diagonalization remains the same. The blocking effects are considered simultaneously. The state energy $ E_{i} $ for each many-particle configuration $|i\rangle$ is
	\begin{eqnarray}   
	\label{eq:E_Pi_alpha}
		\nonumber E_i &=& \sum_{\mu_i\mathrm{(occupied)}}\varepsilon_{\mu_i},
	\end{eqnarray}
       with $\varepsilon_{\mu_i}$ is the energy of the occupied Nilsson state. The $\alpha$ particle energy is obtained as
	\begin{eqnarray}   
	\label{eq:Q_alpha}
		E_\alpha=(E_{p}^\mathrm{P}+E_{p}^\mathrm{N})-(E_{d}^\mathrm{P}+E_{d}^\mathrm{N}),                  
    \end{eqnarray}
       where the superscript P and N denote the state energies for proton and neutron, respectively, and the subscript $p$ and $d$ represent the parent and daughter nuclei, respectively.
\subsection{Empirical formulas for $\alpha$-decay half-lives}
{\label{subsec:formulas-alpha-decay}}
       The Royer formula~\cite{RoyerG2000_JoPGNaPP26_1149} is expressed as 
	\begin{eqnarray}
	    \label{eq:log_T1}
	   	\mathrm{log_{10}} T_{1/2}=a+bA^{1/6}\sqrt{Z} +c\frac{Z}{\sqrt{Q_\alpha}},
        \end{eqnarray}
       where $A$, $Z$, and $Q_\alpha$ represent the mass number, the proton number, and the $\alpha$-decay energy, respectively, while $a$, $b$, $c$, are the adjustable parameters. The Royer formula of Eq.~(\ref{eq:log_T1}) describes the favored $\alpha$ decay well, but when extended to the unfavored $\alpha$ decay, large deviations occur between the calculated $\alpha$ decay half-lives and the experimental data due to the effect of the centrifugal potential. The centrifugal potential leads to a higher total barrier, smaller penetration probability of the barrier, and thus to a longer half-life with respect to the $\alpha$-decay. Therefore, for unfavored $\alpha$ decay, the centrifugal potential plays an important role and cannot be neglected. Based on the original Royer formulas~\cite{RoyerG2000_JoPGNaPP26_1149}, Deng $et~al$. develop the new formula~\cite{Deng2020}, where the effect of angular momentum is considered, which can be expressed as       
	\begin{eqnarray}
	    \label{eq:log_T}
	   	\mathrm{log_{10}} T_{1/2}=a+bA^{1/6}\sqrt{Z} +c\frac{Z}{\sqrt{Q_\alpha}} +dl(l+1)+h.
         \end{eqnarray}
       Here, the fourth term represents the contribution of the centrifugal potential. The angular momentum carried away by the $\alpha$-particle is denoted as $l$. The fifth term represents the blocking effect of unpaired nucleons. For odd-odd nuclei, $a=-26.8125$, $b=-1.1255$, $c=1.6057$, $d=0.0513$, $h=0.7486$ are adopted. 
       
       There are still other formulas developing from the Geiger-Nuttall law~\cite{Ni2008}. It can be expressed as
	\begin{eqnarray}
	    \label{eq:log_T11}
	   	\mathrm{log_{10}} T_{1/2}=a\sqrt{\mu}\frac{Z_1Z_2}{\sqrt{Q_\alpha}}
     +b\sqrt{\mu Z_1Z_2} +c, 
         \end{eqnarray}
       where $Z_1$ and $Z_2$ are the atomic number of daughter and cluster nuclei, respectively. $a=$~0.43464, $b=$~-1.44876, and $c=$~-16.65020 are free parameter coefficient sets obtained by least square fitting to both $\alpha$-decay and cluster radioactivity data. $\mu$ is the reduced mass.
       
       Based on Eq.~(\ref{eq:log_T11}), an improved formula by including the centrifugal potential and the isospin $I=(N-Z)/A$ correlated term proposed in Ref.~\cite{Akrawy2019a} as
	\begin{eqnarray}
	    \label{eq:log_T22}
	   	\mathrm{log_{10}} T_{1/2} &=& a\sqrt{\mu}\frac{Z_1Z_2}{\sqrt{Q_\alpha}}+b\sqrt{\mu Z_1Z_2} +c\\
     &+&dI+eI^2+f[l(l+1)]\notag,
         \end{eqnarray}
       where $a=0.43323$, $b=-1.40527$, $c=-17.13866$, $d=-7.66291$, $e=22.26925$ and $f=0.06902$ are the free parameters obtained by least square fitting to the $\alpha$-decay data. The angular momentum $(l)$ can be calculated by the following equation,
	\begin{equation}\label{eq:Angular momentum}
		 l_{\textrm{min}} =
		\begin{cases}
			\Delta_{j}& \text{for even $\Delta_{j}$ and $\pi_{p}=\pi_{d}$},\\
			\Delta_{j}+1& \text{for even $\Delta_{j}$ and $\pi_{p}\neq\pi_{d}$},\\
			\Delta_{j}& \text{for odd $\Delta_{j}$ and $\pi_{p}\neq\pi_{d}$},\\
			\Delta_{j}+1& \text{for odd $\Delta_{j}$ and $\pi_{p}=\pi_{d}$},
		\end{cases}
	\end{equation}
      where $\Delta_{j}=\rvert j_{p}-j_{d} \rvert$, with $ j_{p} $, $ \pi_{p} $, $ j_{d} $, $ \pi_{d} $ are the spin and parity values of the parent and daughter nuclei, respectively.      

\section{Results and discussions}{\label{Sec:results}}
\subsection{Parameters}

	 \begin{figure*}[htp]
	 	\includegraphics[scale=0.75]{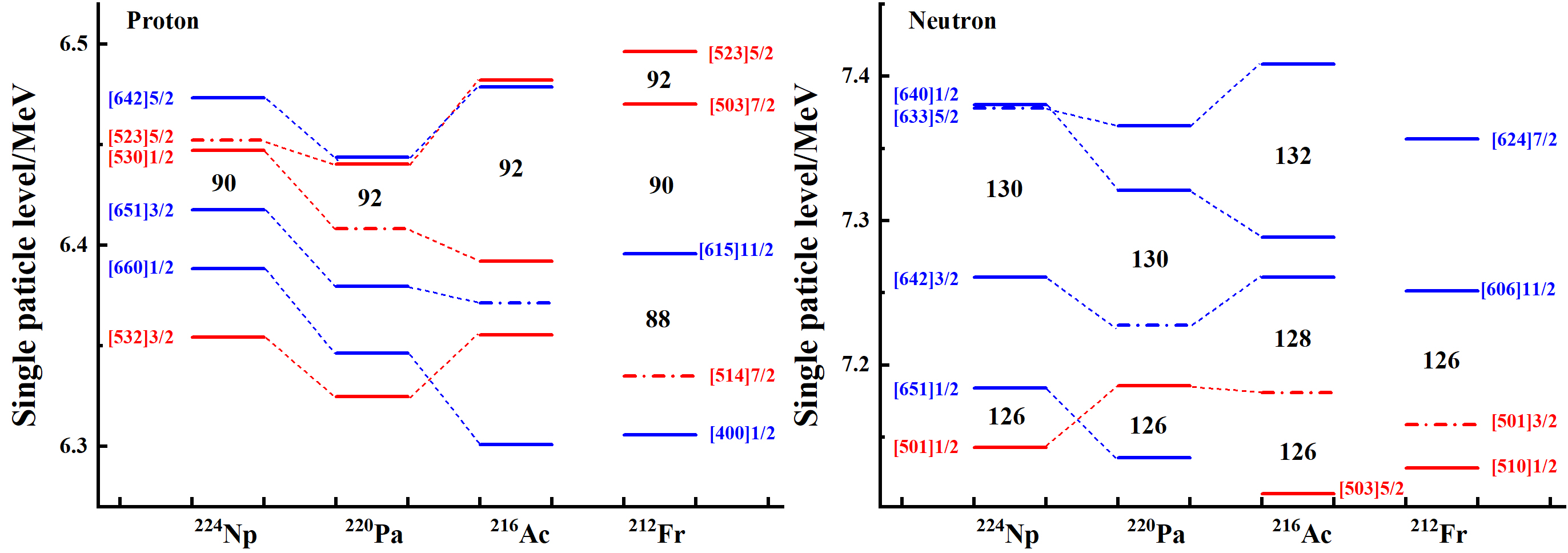} 
	 	\caption{\label{fig:Nilsson}
	 		Single particle levels near the Fermi surface of the nuclei along the $\alpha$-decay chain of $^{224}$Np for protons (left) and neutrons (right). The positive- (negative-) parity levels are denoted by blue (red) lines, and the Fermi surfaces are denoted by dot-dashed lines. The dotted lines are used to guide the eye.}
	 \end{figure*}

	The values of Nilsson parameters $ (\kappa,\mu) $ in Eq.~(\ref{eq:h_Nil}) are taken from the Lund systematics \cite{NilssonS1969_NPA131_1}. The deformations $ \varepsilon_{2} $, $ \varepsilon_{4} $, and $ \varepsilon_{6} $ are input parameters in the PNC calculations. The values used in the present calculations are listed in the Table~\ref{tab:deformation parameters}, where $ \varepsilon_{2} $ and $\varepsilon_{4}$ are larger than the deformation parameters from Ref.~\cite{MoellerP2016_ADaNDT109_1}. In the present calculations, what used to fit the $\alpha $-decay energy is taken from Refs.~\cite{Huang2018,Ma2020}, and the high-order axial deformation $\varepsilon_{6}$ is included due to its important effect on the deformed shell gaps and the further influence on the multiparticle excitation energy.   
	
	\begin{table}[h]
		\centering
		\caption{{\label{tab:deformation parameters}} Deformation parameters $ \varepsilon_{2} $, $ \varepsilon_{4} $, and $ \varepsilon_{6} $ used in the present calculations for $ ^{224}$Np, $ ^{220} $Pa, $ ^{216} $Ac, and $ ^{212} $Fr.}
	\begin{tabular}{p{1.6cm}<{\centering} p{1.6cm}<{\centering} p{1.6cm}<{\centering} p{1.6cm}<{\centering} p{1.6cm}<{\centering}}
		\hline\hline
		 nuclei& $ ^{224} $Np & $ ^{220} $Pa & $ ^{216} $Ac & $ ^{212} $Fr\\
		\hline
            $\varepsilon_2$  &  0.118  &  0.144   &  0.139 & -0.169\\
		$\varepsilon_4$  & -0.021  & -0.018   & -0.030 & -0.027\\
		  $\varepsilon_6$  &  0.004  &  0.004   & -0.038 & -0.010\\
		\hline\hline	
	\end{tabular}	
	\end{table}
	
	The effective pairing strengths $ G_{0} $, $ G_{2} $ can be determined by the odd-even differences in nuclear binding energy in principle. These values are linked with the dimensions of the truncated many-particles configuration space. In the present calculations, the MPC spaces for all the nuclei involved are constructed in the proton $N=4,5,6$ and neutron $N=5,6$ shells. For both of the protons and neutrons, the dimensions of MPC space are about 1300. For all nuclides except $^{224}$Np, the effective monopole- and quadrupole- pairing strengths are $ G_{0}=0.2 $~MeV, $ G_{2}=0.02 $~MeV, respectively, while for $^{224}$Np, $ G_{0}=0.23 $~MeV, $ G_{2}=0.02 $~MeV. The PNC calculations remain stable regardless of the dimension of the MPC space, which has been investigated in Ref. \cite{LiuS2002_PRC66_67301}.

\subsection{Single particle levels}
    The single-particle levels are of importance for determining the low-lying multiparticle states. In the current calculations, Nilsson states are computed for the valence single-particle space including proton $N=4-6$ and neutron $N=5-6$ major shells. 
    
    Figure~\ref{fig:Nilsson} shows the single particle levels near the Fermi surface of nuclei along the $\alpha$-decay chain. The levels are denoted by blue lines with positive parity and red lines with negative parity. Each level is denoted by the quantum numbers $[Nn_z\Lambda]\Omega$. Based on such sequences of single particle levels, the experimental ground state and low-lying excited states can be reproduced well for the neighbor odd-A nuclei, such as the proton ground states in $^{222,223}$Np~\cite{Ma2020,Sun2017}, and $^{230,234}$ Pa ~\cite{KotthausT2013_PRC87_44322,ChuY1978_PRC17_1507} and neutron ground states in $^{223}$U, $^{217}$Th, and $^{213}$Ra \cite{Sun2020,NishioK2000_PRC61_34309}. 
    
    For protons, the Fermi surface changes from $\pi5/2^{-}[523]$ in $^{224}$Np, to $\pi$1/2$^{-}$[530] in $^{220}$Pa, and $\pi$3/2$^{+}$[651] in $^{216}$Ac. The proton $\pi5/2^{-}$[523] level stems from $h_{9/2}$ orbital and $\pi$1/2$^{-}$[530] and $\pi$3/2$^{+}$[651] levels stem from $i_{13/2}$ orbital. The deformed proton shell at $Z=92$ is shown for $^{220}$Pa and $^{216}$Ac while is absent for $^{224}$Np and $^{212}$Fr in Fig.~\ref{fig:Nilsson}. The results of absence of $Z=92$ shell gap in $^{224}$Np is consistent with Ref.~\cite{Sun2017}, where the same trend is observed for $Z<92$, providing no sign of the subshell closure at $Z=92$ from $S_p$ (the single proton separation energies) and $\delta_p$ (separation energy difference). 
    
	\begin{figure*}[htp]
	\includegraphics[scale=0.6]{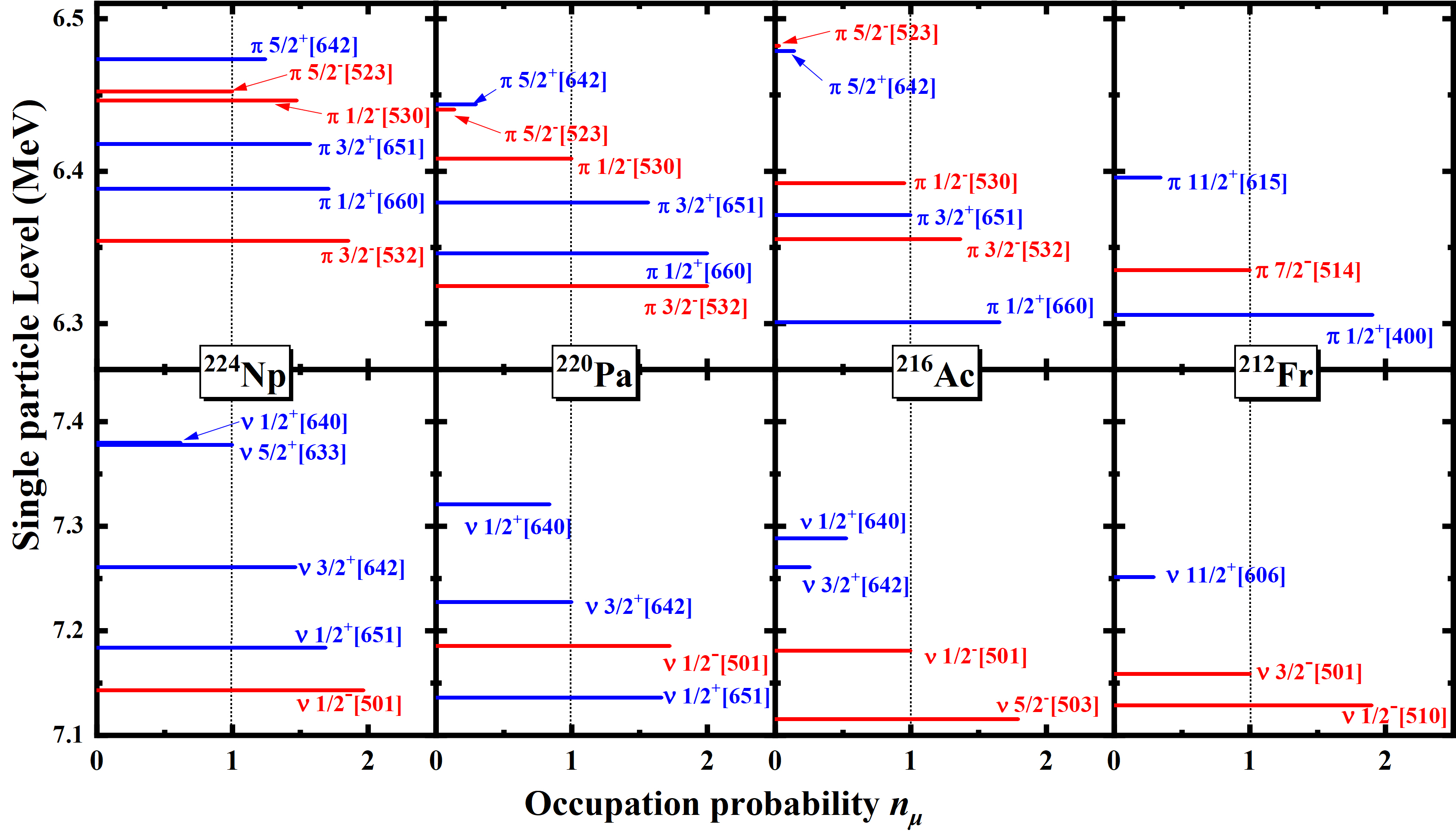} 
	\caption{\label{fig:occupation}
		The occupation probability $ n_{\mu} $ of each proton (top) and neutron (bottom) orbitals $\mu$ near the Fermi surface for the ground states in $ ^{224} $Np, $ ^{220}$Pa, $^{216} $Ac, and $ ^{212} $Fr. The Nilsson levels far below $( n_{\mu}\approx2 )$ and far above $( n_{\mu}\approx0 )$ the Fermi surface are not shown. Positive- (negative-) parity orbitals are denoted by blue (red) lines. The vertical dashed lines are used to guide eye to find the blocked orbitals with $ n_{\mu}=1 $.}
    \end{figure*} 

    The neutron deformed shell gap at $N=130$ is shown in $^{224}$Np and $^{220}$Pa while it disappears in $^{216} $Ac. The neutron Fermi surface locates at $ N = 126 $ neutron shell for $^{212}$Fr. Its deformation is quite different from others, changes from prolate to oblate (see the quadrupole deformation parameters in Table~\ref{tab:deformation parameters}). Therefore, the single-particle level structure differs considerably from these for other nuclei along the $\alpha$-decay chain.

\subsection{Occupation probability and configuration assignment}
	\begin{figure*}[ht]
	 	\includegraphics[scale=0.63]{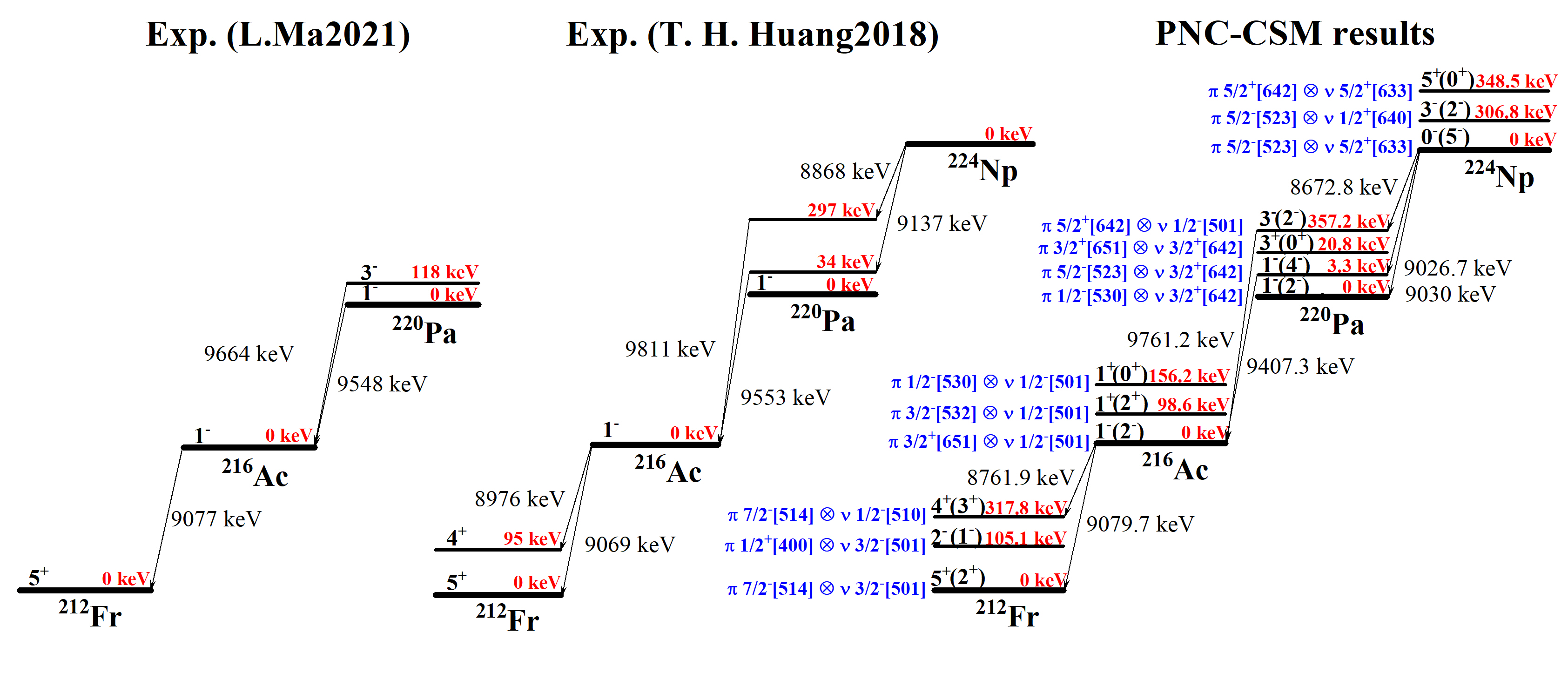} 
	 	\caption{\label{fig:decay stain}
	 		The comparison of the $ \alpha $ particle energy $(E_\alpha)$ and configurations of nuclei in the $\alpha$-decay chain between experimental (Refs.~\cite{Ma2021,Huang2018}) and theoretical (PNC) results. The configurations of the theoretical calculation are on the left of the energy levels. The thick and thin lines represent the ground and low-lying excited states, respectively. The $K^\pi$ values of excited states favored by the GM rules~\cite{GallagherC1962_PR126_1525} are unbracketed. }
  
	 \end{figure*}
  
     The configuration of each multi-particle state is explicitly determined through the wave function. Once the wave function is obtained by Eq. (\ref{eq:eigenstate}), the state configuration can be determined by the occupation probability $ n_{\mu} $, 
	\begin{equation}
		n_{\mu}=\sum_{i}|C_{i}|^{2}P_{i\mu},~~ C_i \text{ is real},
	\end{equation}
     where $P_{i\mu}=1$ represents the state $ |\mu\rangle $ is occupied and $ P_{i\mu} =0$ otherwise. The total particle number $N=\sum_\mu n_\mu$. 
          
     Figure~\ref{fig:occupation} shows the occupation probability $ n_{\mu} $ of each proton(top) and neutron(bottom) orbitals $\mu$ near the Fermi surface at $\hbar\omega$=0 for ground states in $ ^{224} $Np, $ ^{220}$Pa, $^{216} $Ac, and $ ^{212} $Fr, where $\mu$ is double degenerate. An orbital is fully occupied at $ n_{\mu}\approx 2$, and blocked at $ n_{\mu}\approx 1$. The occupation probability of Nilsson levels far below $( n_{\mu}\approx 2)$ and far above $( n_{\mu}\approx0)$ the Fermi surface are not presented. 

     For $^{224}$Np, the blocked orbitals are $ \pi5/2^{-}[523]$ and $\nu5/2^{+}[633] $ with the values of $n_\mu$ of the $\nu5/2^{+}[633] $ orbital and the $ \pi5/2^{-}[523]$ orbital are both $\approx1$. Meanwhile, due to the pairing correlation, others orbitals are partially occupied. In Ref.~\cite{KotthausT2013_PRC87_44322}, $\pi1/2^{-}[530]$ is suggested as the configuration of the ground-state of odd-even Pa isotopes, which is confirmed in the present calculations. From Fig.~\ref{fig:occupation}, the configuration of the ground-state for $^{220}$Pa can be assigned as $\pi1/2^{-}[530]\otimes\nu3/2^{+}[642]$. The ground state configurations of other nuclei can be determined in the same way that is $\pi 3/2^{+}[651]\otimes\nu1/2^{-}[501] $ for $^{216} $Ac, and $ \pi 7/2^{-}[514]\otimes\nu 3/2^{-}[501] $ for $ ^{212} $Fr, respectively. 

     Low-lying excitation states are predicted by the PNC calculations. Although due to the configuration mixing, the occupation probability is not as pure as that for the ground state with $n_{\mu}\approx1$, excitation state configuration can still be determined by the dominant component in their wave function. The excitation energy together with the configuration of the low-lying state for the nuclei along the $\alpha$-decay chain is shown in Fig.~\ref{fig:decay stain}.
     
\subsection{$ \alpha $-decay properties by PNC-DSM and empirical method}
 	
     Figure~\ref{fig:decay stain} illustrates the comparison between experimental and theoretical $\alpha$ particle energies $(E_\alpha)$, as well as the configurations of ground and low-lying excited states. The thick line represents the ground state, while the thin line denotes the low-lying excited state. The theoretically determined configurations are on the left of the energy levels, with the excitation energy above the energy levels. We reproduce the experimentally measured $ \alpha$-decay chain well with $\mathrm{Max}|E_{\alpha}^{\mathrm{cal}}-E_{\alpha}^{\mathrm{exp}}|<300$~keV. Based on the configurations assigned by the PNC method, the minimum angular momentum $l_{min}$ taken away by the $\alpha$ -particle can be obtained by using Eq.~(\ref{eq:Angular momentum}). 
     
     The $\alpha$-decay half-lives are calculated using the empirical formulas Eq.~(\ref{eq:log_T}), Eq.~(\ref{eq:log_T11}), and Eq.~(\ref{eq:log_T22}). The results are listed in the Table~\ref{tab:alpha decay half-lives}, where the first column is the configurations of the ground and low-lying excited states obtained by PNC calculation. The second and third columns show the experimental and calculated $\alpha$-decay energies($Q_\alpha$). The fourth column is the minimum angular momentum. The following three columns display the $\alpha$-decay half-lives calculated by these three different formulas. Subsequently, the available experimental $\alpha$-decay half-lives are listed in the last column. The data for $\alpha$-decay half-lives in experiments are taken from Refs.~\cite{Huang2018,Ma2021}. The deviation between the experimental data and theoretical results lies within two orders of magnitude.

     In the PNC calculation, the ground state of $^{224}$Np is assigned as $ \pi5/2^{-}[523]\otimes\nu5/2^{+}[633]$. The configuration of $0^{-}$ is adopted, which is favored by the Gallagher-Moszkowski (GM) rules~\cite{GallagherC1962_PR126_1525}. If the ground state of $^{224}$Np is adopted as $5^{-}$, in the decay path from the ground state with $ \pi5/2^{-}[523]\otimes\nu5/2^{+}[633]$ in $^{224}$Np to the excited state of $1^{-}$ in $^{220}$Pa, the calculated minimum angular momentum $l_{min}$ will be $4$. Instead of -3.25 and -3.41 with $l_{\min}=2$ by Eq.~(\ref{eq:log_T}) and (\ref{eq:log_T22}), respectively, the calculated $\alpha$-decay half-lives would be -2.02 and -1.75, which far from the experimental result $38\mu s$ $(\rm{log_{10}}T_{1/2}=-4.42)$~\cite{Huang2018}. The angular momentum taken away by the $\alpha$ particle is important. Two calculated low-lying excited states with energy of 306.8 keV and 348.5 keV is assigned as $5^{+}$ $(\pi5/2^{+}[642]\otimes\nu5/2^{+}[633])$ and $3^{-}$ $(\pi5/2^{-}[523]\otimes\nu1/2^{+}[640])$, respectively.
     
     For $^{220}$Pa, its ground state is established as $1^{-}$, which is consistent with the experimental assignment. In the experiment of Ref.~\cite{Huang2018}, $^{220}$Pa exhibits two excited state with energy of 34 keV and 297 keV, respectively, without the detailed configuration information. In the later experimental research~\cite{Ma2021}, an excited state of $3^-$ with 118 keV was observed. The PNC calculations reveal that the first excited state has a configuration of $ \pi5/2^{-}[523]\otimes\nu3/2^{+}[642] $ with excitation energy of 3.3 keV, and a higher excited $3^{-}$ state of $\pi5/2^{+}[642]\otimes\nu1/2^{-}[501]$ with energy of 357.2 keV is predicted. When the excited state $1^{-}$ undergoes favored $\alpha$-decay to the ground state ($1^{-}$) of $^{216}$Ac, the calculated results are -5.24, -5.42, and -5.58, which are compared with the value of -6.51 in Ref.~\cite{Huang2018} in Table~\ref{tab:alpha decay half-lives}. For unfavored $\alpha$-decay of the excited state ($3^{-}$) to the ground state ($1^{-}$) of $^{216}$Ac, the results -5.79, -6.33, and -6.07 compared with the value of -6.63 in Ref.~\cite{Ma2021} in Table~\ref{tab:alpha decay half-lives}. There are three low-lying excited states, which are not shown in Fig.~\ref{fig:decay stain} and Table~\ref{tab:alpha decay half-lives}. That are $ \pi1/2^{-}[530]\otimes\nu1/2^{+}[501] $ (25.1 keV), $\pi5/2^{-}[523]\otimes\nu1/2^{+}[501] $ (335.4 keV), and $\pi3/2^{+}[651]\otimes\nu1/2^{+}[501] $ (352.9 keV) in $^{220}$Pa between $3^+$(20.8 keV) and $3^-$(357.2 keV).
     
     The experimental observed $1^-$ ground state in $^{216}$Ac is assigned as $ \pi 3/2^+[651]\otimes\nu1/2^-[501]$ configuration in the PNC calculations, and two low-lying excited states are predicted at 98.6 keV and 156.2 keV with the configurations of $\pi 3/2^-[532]\otimes\nu 1/2^-[501]$ and $\pi 1/2^-[530]\otimes\nu 1/2^-[501]$, respectively. For $^{212}$Fr, the $5^+$ state is observed in Refs.~\cite{Ma2021,Huang2018}. Its configuration is assigned as $\pi 7/2^-[514]\otimes\nu 3/2^-[501]$ in the present calculation. The $4^+$ excited state is assigned as $\pi 7/2^-[514]\otimes\nu 1/2^-[510]$ with higher energy of 317.8 keV. When the excited state ($1^{-}$) of $^{216}$Ac undergoes unfavored $\alpha$-decay to the ground state ($5^{+}$) of $^{212}$Fr, the calculated $\alpha$-decay half-lives are -3.55, -5.25, -3.35. In this case, the angular momentum taken away by the $\alpha$ particle is as large as 5, which leads to an important effect on the $\alpha$-decay halflive. The results by including the centrifugal terms [Eqs.~(\ref{eq:log_T}) and (\ref{eq:log_T22})] reproduce the experimental better when compared with the value of -3.46 in Ref.~\cite{Ma2021} in Table~\ref{tab:alpha decay half-lives}.

	\section{summary}{\label{Sec:sum}}
	
	In this work, $ ^{224} $Np and its $ \alpha $-decay are investigated by using the particle-number-conserving method in the framework of the deformed shell model and the empirical formulas. The theoretical results of the $\alpha$-decay energy and half-lives for the odd-odd nucleus $ ^{224} $Np and its decay chain can reproduce well of the experimental data. The calculated energy of $\alpha$-particle lies within 300 keV of the experimental data. The Nilsson single-particle levels show the absence of the $Z=92$ subshell for $^{224}$Np. By analyzing the wave function of the deformed shell model Hamiltonian, the configurations are assigned to the experimentally observed ground-state and low-lying excited states. Moreover, more excited states are predicted with the detailed informations of state energy and the configurations by the PNC-DSM. 

    Based on the calculated $\alpha$-decay energy and the assigned configuration of the PNC-DSM, three kinds of the empirical formulas are used to calculate the $\alpha$-decay half-lives. The theoretical results lie in two orders of magnitude errors with comparison of the experimental data. The angular momentum taken away by the $\alpha$ particles is important, especially for the unfavored decay with large angular momentum changing.  
 
       \section*{acknowledgement}{\label{Sec:thanks}}
       This work is supported by the National Key R$\&$D Program of China (Contract Nos.~2024YFE0109804 and 2023YFA1606503), the National Natural Science Foundation of China (Grant No.~12475121) and the China Scholarship Council (Grant No.~EWXM2311280008).
\LTleft=0pt 
\LTright=0pt
\LTcapwidth=174mm
\begin{longtable*}{@{\extracolsep{\fill}}ccccccccc@{}}

    \caption{Calculated $\alpha$-decay half-lives along the decay chains of $^{224}$Np by the empirical formulas based on the PNC calculated $\alpha$-decay energy and configurations. The experimental data are from the Refs.~\cite{Huang2018,Ma2021}.}
    \label{tab:alpha decay half-lives}
   
\endfirsthead
\caption{Calculated $\alpha$-decay half-lives along the decay chains of $^{224}$Np by the empirical formulas based on the PNC calculated $\alpha$-decay energies and configurations. The experimental data are from the Refs.~\cite{Huang2018,Ma2021}.}\\
\hline \hline
         \multicolumn{1}{c}{\multirow{2}{*}{Configuration}} & \multicolumn{2}{c}{$Q_\alpha$/keV}& \multicolumn{1}{c}{\multirow{2}{*}{$l_{min}$}}& \multicolumn{4}{c}{$\lg T_{1/2}$}\\
        \cline{2-3} \cline{5-8}
          & Exp. & Cal. & & Eq.~(\ref{eq:log_T})& Eq.~(\ref{eq:log_T11})& Eq.~(\ref{eq:log_T22})& Exp.\\
          \hline
\endhead
\hline \hline
\endfoot
   \hline \hline
         \multicolumn{1}{c}{\multirow{2}{*}{Configuration}} & \multicolumn{2}{c}{$Q_\alpha$/keV}& \multicolumn{1}{c}{\multirow{2}{*}{$l_{min}$}}& \multicolumn{4}{c}{$\lg T_{1/2}$}\\
        \cline{2-3} \cline{5-8}
          & Exp. & Cal. & & Eq.~(\ref{eq:log_T})& Eq.~(\ref{eq:log_T11})& Eq.~(\ref{eq:log_T22})& Exp.\\ 
        \hline
         $ ^{224} $Np $ \longrightarrow $ $ ^{220} $Pa 
         &        &      &                 &         &    &       &       \\ 
         
         $ \pi5/2^{+}[642]\otimes\nu 5/2^{+}[633] \to \pi 5/2^{+}[642]\otimes\nu 1/2^{-}[501] $
         &      &  9185.3    &  3  & -2.92   & -3.66  & -2.98  &       \\
       
         $ \pi5/2^{+}[642]\otimes\nu 5/2^{+}[633] \to \pi 3/2^{+}[651]\otimes\nu 3/2^{+}[642] $ 
         &      &  9527.8    &  2  & -4.13   & -4.59  & -4.33   &        \\
         $ \pi5/2^{+}[642]\otimes\nu 5/2^{+}[633] \to \pi 5/2^{-}[523]\otimes\nu 3/2^{+}[642] $ 
         &      &  9545.7    & 5  & -2.94  & -4.64  & -2.72  &        \\
         $ \pi5/2^{+}[642]\otimes\nu 5/2^{+}[633] \to \pi 1/2^{-}[530]\otimes\nu 3/2^{+}[642] $ 
         &      &  9549.0    &  5  & -2.95   & -4.65  & -2.73   &        \\
         $ \pi5/2^{-}[523]\otimes\nu 1/2^{+}[640] \to \pi 5/2^{+}[642]\otimes\nu 1/2^{-}[501] $
         &      &  9142.9    &  0  & -3.43   & -3.54  & -3.69  &       \\
         $ \pi5/2^{-}[523]\otimes\nu 1/2^{+}[640] \to \pi 3/2^{+}[651]\otimes\nu 3/2^{+}[642] $ 
         &      &  9485.4    &  1  & -4.22   & -4.48  & -4.49   &        \\
         $ \pi5/2^{-}[523]\otimes\nu 1/2^{+}[640] \to \pi 5/2^{-}[523]\otimes\nu 3/2^{+}[642] $ 
         &      &  9503.2    & 2  & -4.06   & -4.53  & -4.26   &        \\
         $ \pi5/2^{-}[523]\otimes\nu 1/2^{+}[640] \to \pi 1/2^{-}[530]\otimes\nu 3/2^{+}[642] $ 
         &      &  9506.6    &  2  & -4.07   & -4.54  & -4.27   &        \\
         $ \pi5/2^{-}[523]\otimes\nu 5/2^{+}[633] \to \pi 5/2^{+}[642]\otimes\nu 1/2^{-}[501] $
         & 9029.2 \cite{Huang2018}     &  8830.5    &  4  & -1.53   & -2.63  & -1.40  &       \\
         $ \pi5/2^{-}[523]\otimes\nu 5/2^{+}[633] \to \pi 3/2^{+}[651]\otimes\nu 3/2^{+}[642] $ 
         &      &  9173.0    &  3  & -2.89   & -3.62  & -2.95   &        \\
         $ \pi5/2^{-}[523]\otimes\nu 5/2^{+}[633] \to \pi 5/2^{-}[523]\otimes\nu 3/2^{+}[642] $ 
         & 9303.1 \cite{Huang2018}&  9190.8    & 2  & -3.25   & -3.67  & -3.41   & -4.42 \cite{Huang2018}       \\
         $ \pi5/2^{-}[523]\otimes\nu 5/2^{+}[633] \to \pi 1/2^{-}[530]\otimes\nu 3/2^{+}[642] $ 
         &      &  9194.2    &  2  & -3.26   & -3.68  & -3.42   &        \\
         \hline
         $ ^{220} $Pa $ \longrightarrow $ $ ^{216} $Ac\\ 
          $ \pi 5/2^{+}[642]\otimes\nu 1/2^{-}[501] \to \pi 1/2^{-}[530]\otimes\nu 1/2^{-}[501] $ 
         &     &   9782.9   &  3    & -5.11  & -5.94  & -5.26   &        \\
          $ \pi 5/2^{+}[642]\otimes\nu 1/2^{-}[501] \to \pi 3/2^{-}[532]\otimes\nu 1/2^{-}[501] $ 
         &     &   9841.5   &  3    & -5.25  & -6.08  & -5.41   &        \\
          $ \pi 5/2^{+}[642]\otimes\nu 1/2^{-}[501] \to \pi 3/2^{+}[651]\otimes\nu 1/2^{-}[501] $ 
         &  9843.0 \cite{Ma2021}   &   9942.0   &  2    & -5.79  & -6.33  & -6.07   & -6.63 \cite{Ma2021}  \\
         &  9992.7 \cite{Huang2018}&   &    &     &  &  & -7.16 \cite{Huang2018}\\  
          $ \pi 3/2^{+}[651]\otimes\nu 3/2^{+}[642] \to \pi 1/2^{-}[530]\otimes\nu 1/2^{-}[501] $ 
         &     &   9440.2   &  2    & -4.58  & -5.05  & -4.80   &        \\
          $ \pi 3/2^{+}[651]\otimes\nu 3/2^{+}[642] \to \pi 3/2^{-}[532]\otimes\nu 1/2^{-}[501] $ 
         &     &   9498.9   &  2    & -4.73  & -5.21  & -4.95   &        \\
          $ \pi 3/2^{+}[651]\otimes\nu 3/2^{+}[642] \to \pi 3/2^{+}[651]\otimes\nu 1/2^{-}[501] $ 
         &     &   9599.3   &  3    & -4.67  & -5.47  & -4.80   &       \\
          $ \pi 5/2^{-}[523]\otimes\nu 3/2^{+}[642] \to \pi 1/2^{-}[530]\otimes\nu 1/2^{-}[501] $ 
         &     &   9422.4   &  1    & -4.74  & -5.01  & -5.03   &        \\
          $ \pi 5/2^{-}[523]\otimes\nu 3/2^{+}[642] \to \pi 3/2^{-}[532]\otimes\nu 1/2^{-}[501] $ 
         &     &   9481.1  &  1    & -4.89  & -5.16  & -5.18   &        \\
          $ \pi 5/2^{-}[523]\otimes\nu 3/2^{+}[642] \to \pi 3/2^{+}[651]\otimes\nu 1/2^{-}[501] $ 
         &  9729.9 \cite{Huang2018}   &   9581.5   &  0    & -5.24  & -5.42  & -5.58   & -6.51 \cite{Huang2018}  \\
          $ \pi 1/2^{-}[530]\otimes\nu 3/2^{+}[642] \to \pi 1/2^{-}[530]\otimes\nu 1/2^{-}[501] $ 
         &     &   9419.1   &  1    & -4.73  & -5.00  & -5.02   &        \\
          $ \pi 1/2^{-}[530]\otimes\nu 3/2^{+}[642] \to \pi 3/2^{-}[532]\otimes\nu 1/2^{-}[501] $ 
         &     &   9477.7   &  1    & -4.88  & -5.15  & -5.17   &        \\
          $ \pi 1/2^{-}[530]\otimes\nu 3/2^{+}[642] \to \pi 3/2^{+}[651]\otimes\nu 1/2^{-}[501] $ 
         &  9724.8 \cite{Ma2021}   &   9578.1   &  0    & -5.23  & -5.41  & -5.57   & -6.13 \cite{Ma2021}  \\
         \hline
         $ ^{216} $Ac $ \longrightarrow $ $ ^{212} $Fr\\
          $ \pi 1/2^{-}[530]\otimes\nu 1/2^{-}[501] \to \pi 7/2^{-}[514]\otimes\nu 1/2^{-}[510] $ 
         &     &   9806.4   &  4    & -3.64  & -4.80  & -3.60   &        \\
          $ \pi 1/2^{-}[530]\otimes\nu 1/2^{-}[501] \to \pi 1/2^{+}[400]\otimes\nu 3/2^{-}[501] $ 
         &     &   9303.1   &  1     & -5.12  & -5.39  & -5.42   &        \\
          $ \pi 1/2^{-}[530]\otimes\nu 1/2^{-}[501] \to \pi 7/2^{-}[514]\otimes\nu 3/2^{-}[501] $ 
         &     &   9410.2   &  4   & -4.46  & -5.67  & -4.46   &        \\
          $ \pi 3/2^{-}[532]\otimes\nu 1/2^{-}[501] \to \pi 7/2^{-}[514]\otimes\nu 1/2^{-}[510] $ 
         &     &   9027.7   &  4    & -3.48  & -4.64  & -3.44   &        \\
          $ \pi 3/2^{-}[532]\otimes\nu 1/2^{-}[501] \to \pi 1/2^{+}[400]\otimes\nu 3/2^{-}[501] $ 
         &     &   9244.4   &  1    & -4.97  & -5.23  & -5.26   &        \\
          $ \pi 3/2^{-}[532]\otimes\nu 1/2^{-}[501] \to \pi 7/2^{-}[514]\otimes\nu 3/2^{-}[501] $ 
         &     &   9351.5   &  4    & -4.31  & -5.51  & -4.30   &        \\
          $ \pi 3/2^{+}[651]\otimes\nu 1/2^{-}[501] \to \pi 7/2^{-}[514]\otimes\nu 1/2^{-}[510] $ 
         &  9145.4 \cite{Huang2018}   &   8927.2   &  3    & -3.63  & -4.36  & -3.71  &        \\
          $ \pi 3/2^{+}[651]\otimes\nu 1/2^{-}[501] \to \pi 1/2^{+}[400]\otimes\nu 3/2^{-}[501] $ 
         &     &   9143.9   &  2     & -4.51  & -4.96  & -4.72   &        \\
          $ \pi 3/2^{+}[651]\otimes\nu 1/2^{-}[501] \to \pi 7/2^{-}[514]\otimes\nu 3/2^{-}[501] $ 
         &  9240.1 \cite{Huang2018}  &   9251.0   &   5    & -3.55  & -5.25  & -3.35  &    \\
          & 9248.3 \cite{Ma2021}    &      &       &        &   &       & -3.46 \cite{Ma2021}\\

\end{longtable*}    

\newpage
\bibliography{ref}
\end{document}